\documentclass{article}
\usepackage{spconf,amsmath,graphicx,booktabs,url}
\usepackage[hidelinks]{hyperref}
\makeatletter
\renewcommand\subsection{\@startsection{subsection}{2}{\z@}%
  {1.45ex plus .2ex minus .1ex}{0.70ex plus .1ex}{\bfseries}}
\makeatother
\title{Local chord corruption is not recognizer replay:\\structure-matched calibration}
\name{Weiwen Huang \quad Yunda Chen \quad Wangzheng Wu \quad Nengheng Zheng\sthanks{Corresponding author (Email: nhzheng@szu.edu.cn).}}
\address{College of Electronics and Information Engineering, Shenzhen University, China}
\newcommand{\ctarget}{E_{\mathrm{target}}}
\makeatletter
\g@addto@macro\UrlBreaks{\do\/\do-\do.}
\makeatother
\begin{document}
\ninept
\renewcommand{\baselinestretch}{1.00}\normalsize
\maketitle

\begin{abstract}
Synthetic chord substitutions offer controlled tests of music generation, but their effects can differ from those of a complete recognized chord sequence. We propose structure-matched calibration, which constructs synthetic chord sequences that match the changed positions and harmonic-relation composition of recognizer replay. Paired generation compares both target-response magnitude and output-chord agreement with replay. On 29 of 30 MUSDB18-HQ songs, central four-second tritone corruption produces a larger target response than complete replay in MIDI-SAG. On 24 held-out MoisesDB songs, structure matching reduces target-response distance to replay by 81\% for MIDI-SAG and 77\% for MusicGen-Chord. Distance decreases on every song in both models with CNN--CRF. Joint matching also reduces output-chord mismatch with replay by 8--17 percentage points relative to temporal or relational matching alone.
\end{abstract}
\section{Introduction}
Singing accompaniment generation (SAG) creates instrumental backing for a vocal performance. Recent systems such as MIDI-SAG \cite{tsai2026midisag} and MusicGen-Chord \cite{jung2024musicgenchord,copet2023musicgen} accept chord sequences as explicit conditioning inputs, making harmony a central control interface. Other SAG systems generate accompaniment directly from singing or target faster non-autoregressive synthesis \cite{donahue2023singsong,chen2024fastsag}, while controllable music models expose time-varying controls \cite{wu2024musiccontrolnet}. When an existing recording provides the harmonic guide, automatic chord recognition (ACR) can estimate these sequences from the mixture. Differences in root, quality, and no-chord events then propagate through the chord input to the generated accompaniment. Understanding this propagation is essential for evaluating chord-conditioned systems.\footnote{Code and derived results: \url{https://github.com/Viwennnnnn/local-chord-corruption-replay}}

Prior work has used controlled chord substitutions to study generator sensitivity \cite{gao2024chordconditioned}; Accompaniment Prompt Adherence (APA) evaluates prompt adherence with synthetic perturbations and listening judgments \cite{grachten2025apa}, while COCOLA uses learned audio representations to measure harmonic and rhythmic coherence between stems \cite{ciranni2025cocola}. These measures assess accompaniment adherence or coherence, whereas we ask whether a synthetic chord probe reproduces the downstream effect of a complete recognized sequence. Chord-recognition research has studied label vocabularies and timing boundaries \cite{harte2005symbolic,pauwels2013evaluating,chen2019harmonytransformer}, harmonic relations and pitch-space geometry \cite{carsault2018musical,kinnaird2021hierarchy,humphrey2012tonnetz}, and broader ACR evaluation \cite{mcvicar2014review,pauwels2019twenty}.

We distinguish two questions: whether a generator responds to chord changes, and whether a synthetic probe reproduces its response to complete recognizer replay. A local tritone probe concentrates one harmonic relation into a short interval. Complete recognition instead produces a sequence of root, quality and no-chord changes distributed across the excerpt. These differences make representativeness an empirical question, separate from sensitivity to an isolated substitution. A matched comparison must therefore examine both the strength of the response and the resulting harmonic sequence. We test whether preserving the temporal support and harmonic-relation composition of replay makes synthetic probes more representative.

\begin{figure*}[t]
\centering
\includegraphics[width=\textwidth]{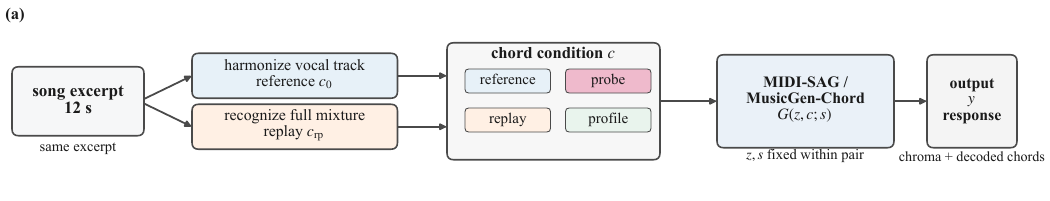}\\[-1pt]
\includegraphics[width=\textwidth]{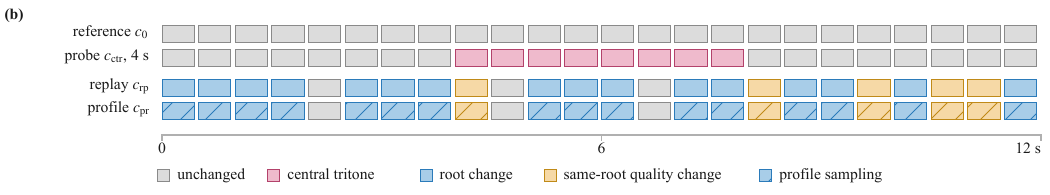}
\caption{Paired evaluation of four chord conditions. (a) A song excerpt provides two automatically obtained chord sources: vocal-track harmonization gives the reference and full-mixture ACR gives replay. Within a pair, the generator, non-chord inputs and seed are fixed, so only the chord condition changes. (b) MoisesDB construction on the shared 24~$\times$~0.5-s grid. Replay changes the cells shown in blue or yellow; the central probe changes eight cells, and the profile retains replay's temporal support and per-cell relation categories while sampling new labels.}
\label{fig:conditions}
\end{figure*}

We address this through \emph{structure-matched calibration} (Fig.~\ref{fig:conditions}): synthetic chord sequences that match replay's changed positions and harmonic-relation composition while resampling chord labels. The findings hold across three generators (MIDI-SAG \cite{tsai2026midisag}, MusicGen-Chord \cite{jung2024musicgenchord}, AccoMontage \cite{zhao2021accomontage}), two recognizers---a convolutional neural network with a conditional random field (CNN--CRF) \cite{korzeniowski2016fullyconv} and DeepChroma with a conditional random field (DeepChroma+CRF) \cite{korzeniowski2016deepchroma,bock2016madmom}---and two datasets (MUSDB18-HQ \cite{rafii2019musdb18hq}, MoisesDB \cite{pereira2023moisesdb}).

Local corruption and complete replay are not interchangeable: in MIDI-SAG, a four-second central tritone probe produces a larger target response on 29 of 30 MUSDB18-HQ songs. Replay's divergence from the reference spans 10.73 s on average, more than twice the probe duration. Matching the structure of that divergence cuts the response distance to replay by 81\% for MIDI-SAG and 77\% for MusicGen-Chord on 24 held-out MoisesDB songs under CNN--CRF, with the distance falling on every song in both models. Partial matching already approximates replay's response magnitude, while joint temporal and relational matching further improves agreement with chords decoded from replay-generated audio. Response magnitude and output-chord agreement must therefore be reported together. Section~\ref{sec:method} defines the conditions and metrics, Section~\ref{sec:results} reports the paired comparisons and ablation, and Section~\ref{sec:conclusion} concludes.

\section{Method}\label{sec:method}
\subsection{Experimental setup}
Let $G(z,c;s)$ generate music from non-chord inputs $z$ and chord sequence $c$ under seed $s$; depending on the model, $z$ contains vocals, melody, context, or a text prompt. Chord conditions are obtained automatically along two routes: vocal-track harmonization yields the reference sequence $c_0$ using the AccoMontage2 harmonization pipeline \cite{yi2022accomontage2}, while full-mixture recognition yields replay sequence $c_{\mathrm{rp}}$. We define the baseline output as $y_0=G(z,c_0;s)$ and study the deviation introduced by $c_{\mathrm{rp}}$. A difference between replay and $c_0$ is therefore a deviation from the reference, not a recognition error against ground truth. Within a pair the generator, its non-chord inputs and the seed are all held fixed, so the chord condition is the only variable (Fig.~\ref{fig:conditions}a).

We compare four conditions. \emph{Baseline} is $c_0$. \emph{Replay} $c_{\mathrm{rp}}$ is the full-mixture ACR prediction read at the same scoring window. \emph{Central corruption} $c_{\mathrm{ctr}}$ shifts chord roots by six semitones over the central four seconds, leaving no-chord labels (N) unchanged. \emph{Profile} $c_{\mathrm{pr}}$ matches replay's changed positions and harmonic relations while retaining baseline labels elsewhere. Profile labels are sampled within the matched categories and may coincide with replay when a category admits no alternative.

The three generators expose different interfaces: MIDI-SAG is an audio diffusion model driven by melody and chord controls, MusicGen-Chord is an audio model receiving a chord progression as a half-second control sequence, and AccoMontage is symbolic and beat-based, arranging accompaniment by phrase selection and reharmonization. Two recognizers are used so that no conclusion depends on one recognizer: CNN--CRF \cite{korzeniowski2016fullyconv} and DeepChroma+CRF \cite{korzeniowski2016deepchroma,bock2016madmom}, denoted CNN and Deep in tables.

\subsection{Profile construction}
Temporal support is the set of scoring blocks whose replay labels differ from baseline (Fig.~\ref{fig:conditions}b). Each baseline--replay pair is assigned a relation category---agreement, same-root quality, relative substitution, semitone-root, tritone-root, fifth-root, other-root, or no-chord-involved---over a vocabulary of major and minor triads plus no-chord. Both dataset constructions preserve the changed positions and retain baseline labels elsewhere. On MUSDB18-HQ, the profile matches relation-category counts across changed cells; on MoisesDB, it also preserves the relation category at each changed cell. We sample labels compatible with the assigned categories, preferring alternatives to replay wherever the category permits. A fixed per-song seed selects the profile before generation. Both constructions match temporal support and relation composition; MoisesDB additionally matches the relation category at each support cell.

\subsection{Evaluation metrics}
For each half-second scoring block, $x_t$ is the normalized baseline-output chroma vector and $y_t$ is the normalized altered-output chroma vector. The terms $h_0(t)$ and $h_1(t)$ are the chord templates implied by the baseline and altered conditions. Over the blocks $\mathcal I$ that changed relative to baseline, the target response is
\begin{equation}
\begin{split}
\ctarget={}&\operatorname*{mean}_{t\in\mathcal I}\{\cos(y_t,h_1)-\cos(y_t,h_0)\\
&\qquad-\cos(x_t,h_1)+\cos(x_t,h_0)\},
\end{split}
\end{equation}
where $\cos(\cdot,\cdot)$ is cosine similarity. A positive value means the altered output moves toward its own chord target relative to baseline output. This difference of relative similarities can exceed one. Across conditions, $\ctarget$ compares response magnitude along each condition's prescribed harmonic change; output-chord agreement is measured separately below. We also measure output chroma distance from baseline over $\mathcal I$ and over the full window.

Averaging $\ctarget$ over three generation seeds within a song gives $\bar E$. Calibration gain compares how close each synthetic condition lands to replay:
\begin{equation}
\Delta_{\mathrm{profile}}=|\bar E_{\mathrm{central}}-\bar E_{\mathrm{replay}}|
-|\bar E_{\mathrm{profile}}-\bar E_{\mathrm{replay}}|,
\end{equation}
so $\Delta_{\mathrm{profile}}>0$ favors the profile. The primary measure uses chroma energy normalized statistics (CENS) \cite{muller2011chromatoolbox}, a standard representation of harmonic content that emphasizes harmony and is stable to timbre and dynamics; constant-Q transform (CQT) chroma \cite{mcfee2015librosa} is a second standard representation with different time--frequency resolution, so agreement between the two cannot be attributed to one front end.

For the primary replay path, CNN--CRF supplies $c_{\mathrm{rp}}$. We decode every generated waveform, including replay, with DeepChroma+CRF on the same 24-cell grid. Output-chord mismatch compares these generated sequences, not input chord labels.

\subsection{Experimental protocol}
Every excerpt is scored on a uniform $24\times0.5$~s grid, using one 12-s excerpt from each of 30 MUSDB18-HQ songs and 24 independent MoisesDB songs \cite{rafii2019musdb18hq,pereira2023moisesdb}. MoisesDB selection balances genres, allows one song per artist group, and is driven by vocal activity only; the 24 songs were fixed before a separate six-song pilot, so no selection step could consult generated output. MIDI-SAG and MusicGen-Chord share the same 24 half-second chord labels and use three paired seeds per condition, and MIDI-SAG retains its 47.55-s context. AccoMontage receives a 48-beat sequence with donor phrases fixed across conditions; because its output is symbolic and beat-based rather than a half-second audio-control grid, its response is measured by projecting the output pitch-class change onto the altered chord template.

Ablations separate the two matching factors: temporal-only matching shifts roots up seven semitones at replay's changed positions while preserving quality, relation-only matching places replay's relation proportions in the central four seconds, and joint matching preserves both. We also test a second recognizer and three independently sampled profiles, averaging their individual replay distances.

Songs, not seeds, are the statistical units. We use 10,000 song-bootstrap resamples for 95\% confidence intervals (CIs) and two-sided paired signed-rank tests at $\alpha=0.05$. The two primary model comparisons form one Holm family; recognizer, profile, and ablation comparisons are corrected within their own families; the six MUSDB tests use Benjamini--Hochberg correction.

\section{Results}\label{sec:results}
\subsection{Local corruption and recognizer replay}
\begin{figure}[t]
\centering
\includegraphics[width=\columnwidth]{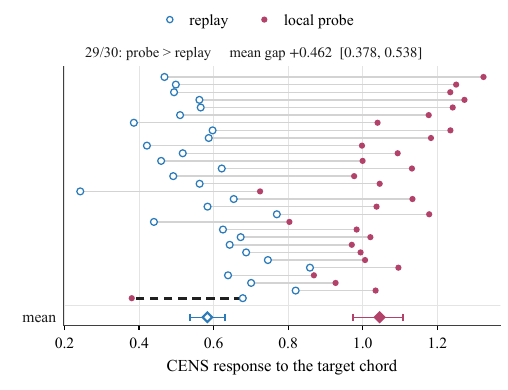}
\caption{MIDI-SAG: local probe versus CNN--CRF replay on 30 MUSDB18-HQ songs, sorted by probe-minus-replay target response. Each row is a three-seed song mean; bottom markers show condition means (blue: replay; red: local probe) with 95\% song-bootstrap intervals. The top annotation gives the mean paired gap and its 95\% interval. The black dashed connector marks the single reversal where replay exceeds the local probe.}
\label{fig:paired}
\end{figure}

In MIDI-SAG on MUSDB18-HQ, central tritone corruption produced a larger target response than complete CNN--CRF replay (Fig.~\ref{fig:paired}). The CENS gap was positive on 29/30 songs, with a mean of 0.462 and 95\% CI $[0.378,0.538]$. The difference was significant on a two-sided paired signed-rank test (BH-adjusted $p=7.82\times10^{-8}$). Output chroma distance from baseline over changed regions was larger on 28/30 songs, with a mean gap of 0.146 (BH-adjusted $p=1.14\times10^{-6}$). The mean target gap stayed between 0.450 and 0.469 across seeds. Divergence from the reference occupied 10.73 s on average for replay, compared with the four-second central probe. That divergence was dominated by root changes (76.12\% of its event mass), with tritone substitutions at only 1.79\%. In a separate relation-diagnostic analysis of 18 songs from this cohort, relative substitutions produced 2.88 times the CENS full-window output change of same-root quality flips, so comparable divergence need not induce comparable responses.

\begin{figure}[t]
\centering
\includegraphics[width=0.90\columnwidth]{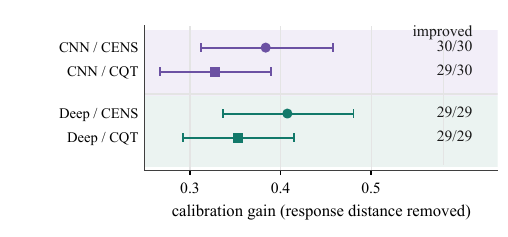}
\caption{MIDI-SAG profile calibration gain relative to central corruption on MUSDB18-HQ. CNN and Deep denote CNN--CRF and DeepChroma+CRF. Points show mean gains in target-response distance to replay; bars are 95\% song-bootstrap intervals. Shaded rows group the recognizer (purple: CNN--CRF; green: DeepChroma+CRF), circles and squares denote CENS and CQT, respectively, and the right column reports songs with positive gain. Deep uses 29 songs because one profile admitted no non-identical synthetic target.}
\label{fig:calib}
\end{figure}

For MIDI-SAG with CNN--CRF on MUSDB18-HQ, structure matching reduced mean CENS target-response distance to replay from 0.482 for central corruption to 0.098 for the profile. Their difference, 0.384, is the calibration gain plotted in Fig.~\ref{fig:calib}. The profile's lower replay distance also holds with a second recognizer and a second chroma representation. We next tested calibration on a new multitrack source and a third architecture.

\subsection{Cross-model results}
Structure matching reduced CENS response distance by approximately 81\% for MIDI-SAG and 77\% for MusicGen-Chord on MoisesDB (Table~\ref{tab:crossed}), improving all 24 songs in both models under CNN--CRF. Mean CENS gains were 0.369 and 0.337, with 95\% CIs $[0.299,0.443]$ and $[0.263,0.414]$; both comparisons remained significant after Holm correction (adjusted $p\leq2.38\times10^{-7}$). Secondary CQT distance fell from 0.386 to 0.071 for MIDI-SAG and from 0.367 to 0.083 for MusicGen-Chord; these means are included in the released metric tables. Replacing CNN--CRF with DeepChroma+CRF preserved the CENS improvement (adjusted $p\leq7.15\times10^{-7}$), as did averaging replay distances over three independently sampled profiles (24/24 MIDI-SAG, 23/24 MusicGen-Chord).

\begin{table}[t]
\caption{MoisesDB calibration: mean CENS target-response distance to replay (24 songs; lower is better). Improved counts songs where Profile is closer than Central. CNN denotes CNN--CRF; Deep denotes DeepChroma+CRF.}
\label{tab:crossed}
\par\vspace{7pt}
\centering
\setlength{\tabcolsep}{4pt}
\begin{tabular}{ccccc}
\toprule
ACR & Generator & Central & Profile & Improved\\
\midrule
CNN & MIDI-SAG & 0.4570 & \textbf{0.0877} & 24/24\\
CNN & MusicGen-Chord & 0.4361 & \textbf{0.0990} & 24/24\\
Deep & MIDI-SAG & 0.4488 & \textbf{0.0782} & 23/24\\
Deep & MusicGen-Chord & 0.4492 & \textbf{0.1110} & 23/24\\
\bottomrule
\end{tabular}
\end{table}

Direct output-chord comparisons showed the same direction of improvement under CNN--CRF replay (Fig.~\ref{fig:ablation}b). From Central to Profile, full-window mismatch fell from 86.11\% to 66.20\% in MIDI-SAG and from 90.16\% to 76.85\% in MusicGen-Chord. Thus, the reduction in target-response distance was accompanied by closer agreement between the generated chord sequences.

Calibration also transferred to AccoMontage's native beat-based interface, where target-response distance fell by approximately 74\% (Table~\ref{tab:native}), with positive gain on 23/24 songs; the target-response and raw pitch-class tests gave $p=2.98\times10^{-6}$ and $p=2.47\times10^{-5}$, respectively. The one song without improvement is consistent with relation categories that leave little room to vary chord identity once mapped onto beats. This third generator reproduces the effect in a symbolic, beat-normalized representation rather than a half-second audio-control grid.

\begin{table}[t]
\caption{AccoMontage calibration on 24 songs. Gain is Central's distance to replay minus Profile's; Improved counts positive song-level gains. Intervals are song-bootstrap 95\% CIs.}
\label{tab:native}
\par\vspace{7pt}
\centering
\setlength{\tabcolsep}{5pt}
\begin{tabular}{ccc}
\toprule
Measure & Gain [95\% CI] & Improved\\
\midrule
Target response & 0.272 [0.193, 0.346] & 23/24\\
Pitch-class distance & 0.240 [0.154, 0.326] & 21/24\\
\bottomrule
\end{tabular}
\end{table}

\subsection{Ablation study}
\begin{figure}[t]
\centering
\includegraphics[width=0.88\columnwidth]{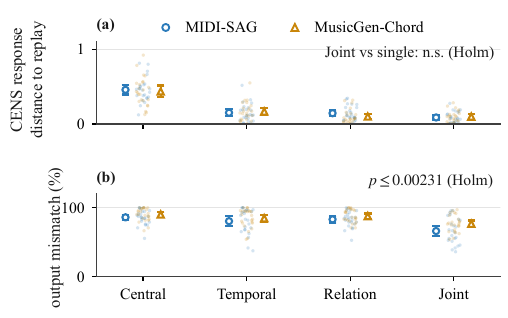}
\caption{Ablation on 24 MoisesDB songs. (a) CENS distance to replay; (b) decoded output-chord mismatch with replay-generated audio. Temporal and Relation match one factor; Joint matches both. Faint markers are song means across seeds; symbols and bars show means and 95\% bootstrap intervals. Annotations give Holm-corrected Joint-versus-single-factor tests.}
\label{fig:ablation}
\end{figure}

Both single-factor conditions reduced mean CENS distance, and the full profile achieved the lowest mean (Fig.~\ref{fig:ablation}a), but none of the four full-versus-single-factor CENS comparisons was significant after Holm correction. The two metrics answer different questions: single factors can approximate response magnitude, while output-chord agreement tests reproduction of replay's harmonic sequence. Central and temporal-only matching differ in both position and substitution relation, so their gap is not a pure position effect. Output chord sequences revealed an additional benefit of joint matching.

Relative to the two single-factor conditions, joint matching reduced output-chord mismatch with replay (Fig.~\ref{fig:ablation}b) by 14.35--16.96 percentage points in MIDI-SAG and 8.04--11.05 in MusicGen-Chord. All four comparisons remained significant after Holm correction ($p\leq0.00231$) and favored the full profile on 17 to 21 of 24 songs. Single-factor conditions therefore approximate response magnitude without matching the generated chord sequence as closely.

\subsection{Discussion}
Two objectives must be reported separately: reproducing the strength of replay's response, and reproducing its harmonic consequences. A small target-response distance is insufficient because a generator that ignores chord changes yields small distances under every condition; output-chord agreement tests whether the synthetic condition also induces replay's harmonic sequence. The temporal grid must match the generator's representation, hence the beat mapping for AccoMontage. In the separate six-song timing pilot noted in Section~2.4, shifting replay support by 0.5 s earlier or later yielded improvement on only four songs in either direction. Replay fidelity is therefore sensitive to temporal alignment and remains untested in generators without explicit chord controls.

The results point to a practical principle: synthetic perturbations should match the control interface through which a system receives harmony, so recognizer-aware benchmarks can separate local sensitivity from replay fidelity across generators and control interfaces. Because profiles are generated from recognizer outputs without copying their labels, the protocol can scale to new songs, recognizers and control granularities; learning profile distributions across genres and testing structural agreement alongside perceived accompaniment quality offers a direct route toward a practical recognizer-in-the-loop benchmark.

\section{Conclusion}\label{sec:conclusion}
Local corruption and recognizer replay answer different questions, and structure-matched profiles narrow their gap across MIDI-SAG, MusicGen-Chord, and AccoMontage. On 24 MoisesDB songs, both audio models improve on every song; repeated profiles and a second recognizer preserve the effect, while joint matching best reproduces replay-generated chords. Recognized harmony should therefore be evaluated as a sequence, reporting response magnitude and output structure together.

\clearpage
\small
\makeatletter
\let\icassp@thebibliography\thebibliography
\let\icassp@endthebibliography\endthebibliography
\renewenvironment{thebibliography}[1]{%
  \icassp@thebibliography{#1}%
  \setlength{\itemsep}{-0.8pt}%
  \setlength{\parsep}{0pt}%
}{\icassp@endthebibliography}
\makeatother
\bibliographystyle{IEEEbib}
\bibliography{references}
\end{document}